\documentclass[preprint,12pt,3p]{elsarticle}

\usepackage{amssymb}
\usepackage{graphicx}
\usepackage{mathrsfs}
\usepackage{amsmath}
\usepackage{ulem}

\usepackage{color, soul}

\renewcommand\labelenumi{(\roman{enumi})}
\renewcommand\theenumi\labelenumi

\journal{Computational Condensed Matter}

\begin{document}

\begin{frontmatter}

\title{Bloch oscillations in graphene from an artificial neural network study}

\author[label1]{M Carrillo}
\address[label1]{Laboratorio de Inteligencia Artificial y Superc\'omputo, Instituto de F\'isica y Matem\'aticas, Universidad Michoacana de San Nicol\'as de Hidalgo, Morelia, 58040, M\'exico}
\ead{mcarrillo@ifm.umich.mx}

\author[label1]{J A Gonz\'alez}
\ead{gonzalez@ifm.umich.mx}

\author[label2]{S Hern\'andez-Ortiz}
\address[label2]{Instituto de Ciencias Nucleares, Universidad Nacional Aut\'onoma de M\'exico, Apartado  Postal 70-543, Ciudad de M\'exico, 04510, M\'exico.}
\ead{saul.hernandez@correo.ciencias.unam.mx}

\author[label1]{C E L\'opez}
\ead{clopez@ifm.umich.mx}

\author[label1]{A Raya}
\ead{raya@ifm.umich.mx}

\begin{abstract}
We develop an artificial neural network (ANN) approach to classify simulated signals corresponding to the semi-classical description of Bloch oscillations in pristine graphene. After the ANN 
is properly trained, we consider the inverse problem of Bloch oscillations (BO), namely, 
 a new signal is classified according to the external
electric field strength oriented along either the zig-zag or arm-chair edges of the graphene membrane, with a correct classification that ranges from 82.6\% to 99.3\% depending on the accuracy of the predicted electric field. This approach can be improved depending on the time spent in training the network
and the computational power available. Findings in this work can be straightforwardly extended to a 
variety of Dirac-Weyl materials.
\end{abstract}


\end{frontmatter}

\section{Introduction}
Even though the band structure of graphene has been known for 70 years from the seminal work of Wallace~\cite{Wal}, it was
soon after the first isolation of its membranes~\cite{graphene1,graphene2,graphene3} that material science has underwent a continuum revolution toward the era of two-dimensional Dirac-Weyl materials~\cite{Weyl}. Novel properties of the collective excitations of these materials, namely, ultrarelativistic fermions, allow to establish direct connection with fundamental physics in particle physics colliders. Nevertheless, some traditional solid state effects are still of relevance to explore in graphene-like materials, such as the phenomenon of Bloch oscillations (BO)~\cite{Bloch,Zener}. Although BO are not observed in real solids, they represent a favorite example of the influence of a periodic array (and an external force field) in the quantum motion of charge carriers. Actual BO are directly
observed under a variety of experimental conditions in high-purity semiconductor
superlattices~\cite{exp1,exp2,exp21,exp22,exp3,exp31,exp4,exp41,exp5,exp51} as well as other systems with similar properties to bulk crystals, including atomic systems~\cite{atom1,atom2}, dielectric~\cite{dielec1,dielec2,dielec3}, plasmonic waveguide arrays~\cite{plas} and bilayer graphene superlattices~\cite{cheng,changan}. Therefore, there is an obvious relevance in the study of such systems that goes beyond solids. The inverse problem of BO has already been addressed by our group for crystal structures in one and two dimensions,  categorizing simulated signals of BO via an Artificial Neural Networks (ANN) approach \cite{IBP,els}. These works give confidence enough to implement ANNs in more complex systems such as the one reviewed here, namely a Dirac-Weyl material.

\section{Bloch oscillation in Graphene: Semiclassical approach}\label{sec:BO}

\begin{figure}
\centering
	\includegraphics[width=0.68\textwidth]{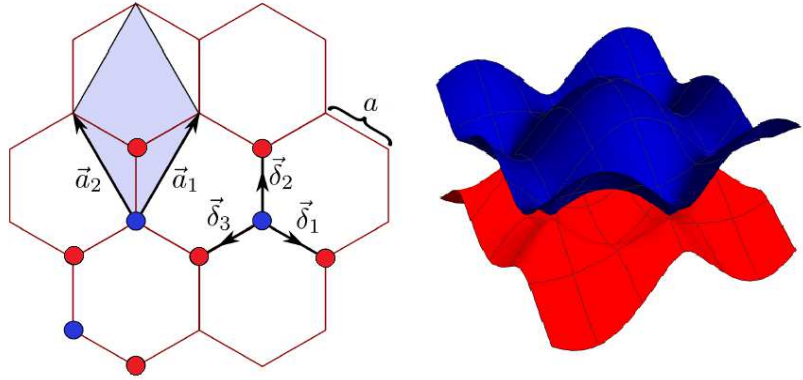}
	\caption{Left: Crystal structure of graphene. Primitive vectors $\mathbf{a}_{1,2}$ and  vectors connecting each atom to its nearest neighbors $\boldsymbol{\delta}_{1,2,3}$ are shown. Right: Energy-momentum relation in the first Brillouin zone. Linear behavior is found near the so-called Dirac points.}
	\label{fig:crystal}
\end{figure}

The crystal structure of graphene (and other Dirac-Weyl materials) consists of a one-atom thick array of carbon atoms (or other) tightly packed in a honeycomb lattice as shown in Fig.~\ref{fig:crystal}. In real space, the hexagonal array is better described in terms of two triangular sublattice with primitive vectors
\begin{equation}
\mathbf{a}_1=\frac{\sqrt{3}}{2}a \hat{e}_x+\frac{3}{2}a\hat{e}_y\;, \qquad
\mathbf{a}_2=-\frac{\sqrt{3}}{2}a \hat{e}_x+\frac{3}{2}a\hat{e}_y\;,
\end{equation}
where $\hat{e}_{x,y}$ are unit vectors along of the graphene membrane in the plane, with an interatomic distance $a\simeq 1.42$ \textup{\AA}.
Each atom of a given sublattice is connected to its nearest neighbors, which in turn belong to other sublattices, through the vectors
\begin{eqnarray}
\boldsymbol{\delta}_1=\frac{\sqrt{3}}{2}a \hat{e}_x+\frac{a}{2}\hat{e}_y
\;,\qquad  
\boldsymbol{\delta}_2=-\frac{\sqrt{3}}{2}a \hat{e}_x+\frac{a}{2}\hat{e}_y
\;,\qquad 
\boldsymbol{\delta}_3=-a\hat{e}_y\;.
\end{eqnarray}
Correspondingly, the tight-binding Hamiltonian is of the form
\begin{equation}
H(\mathbf{k})=\left( \begin{array}{cc} 0 & \tau \varepsilon(\mathbf{k}) \\ \tau \varepsilon^*(\mathbf{k}) & 0 \end{array}\right)\;,
\end{equation}
where $\mathbf{k}=(k_x,k_y)$ is the electron crystal momentum vector, $\tau$ the hopping parameter and
\begin{equation}
\varepsilon(\mathbf{k})=\sum_{\boldsymbol{\delta}}e^{i\mathbf{k}\cdot \boldsymbol{\delta}} = 2ie^{\frac{i}{2}k_x a}\sin\left(\frac{\sqrt{3}}{2} ak_x \right)+e^{-iak_x}\;.
\end{equation}
Therefore, the energy-momentum dispersion relation is $\epsilon(\mathbf{k})= \pm \tau|\varepsilon(\mathbf{k})|,$ which can be written conveniently as
\begin{equation}\label{disprel}
\varepsilon(\mathbf{k})=\sqrt{5+4\cos{\left(\frac{3}{2}ak_x \right)}\cos{\left(\frac{\sqrt{3}}{2}ak_y \right)} - 4\sin^2\left( \frac{\sqrt{3}}{2}ak_y\right)}\;.
\end{equation}

As explained below, many features of Bloch oscillations (BO) in graphene and other Dirac-Weyl semi-metals with an underlying hexagonal lattice, such as the one shown in Fig.~\ref{fig:crystal}, can be derived from this dispersion relation.

\begin{figure}
	\centering
	\includegraphics[width=12cm]{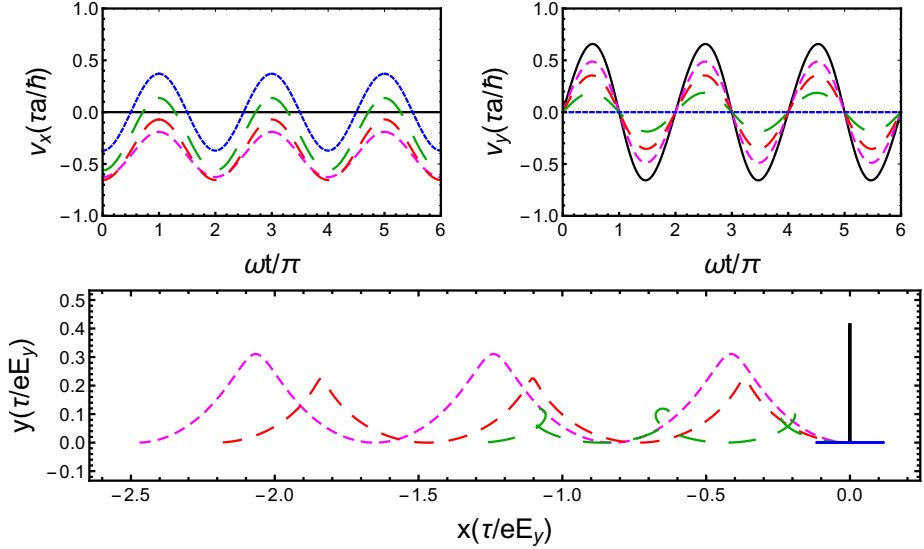}
	\caption{The time dependence of $v_x$, $v_y$ and the trajectories of the electron on the graphene sheet for different $k_x$ at $\frac{3}{2}ak_y = 0$ (solid dark line), $\frac{3}{2}ak_y = \pi/4$ (dash magenta line), $\frac{3}{2}ak_y = \pi/3$ (dash red line), $\frac{3}{2}ak_y = 5\pi/12$ (dot green line), and $\frac{3}{2}ak_y = \pi/2$ (solid blue line). The electric field $\mathbf{E}$ is along the $y$ direction. }
	\label{FULLEy}
\end{figure}

\begin{figure}
	\centering
	\includegraphics[width=12cm]{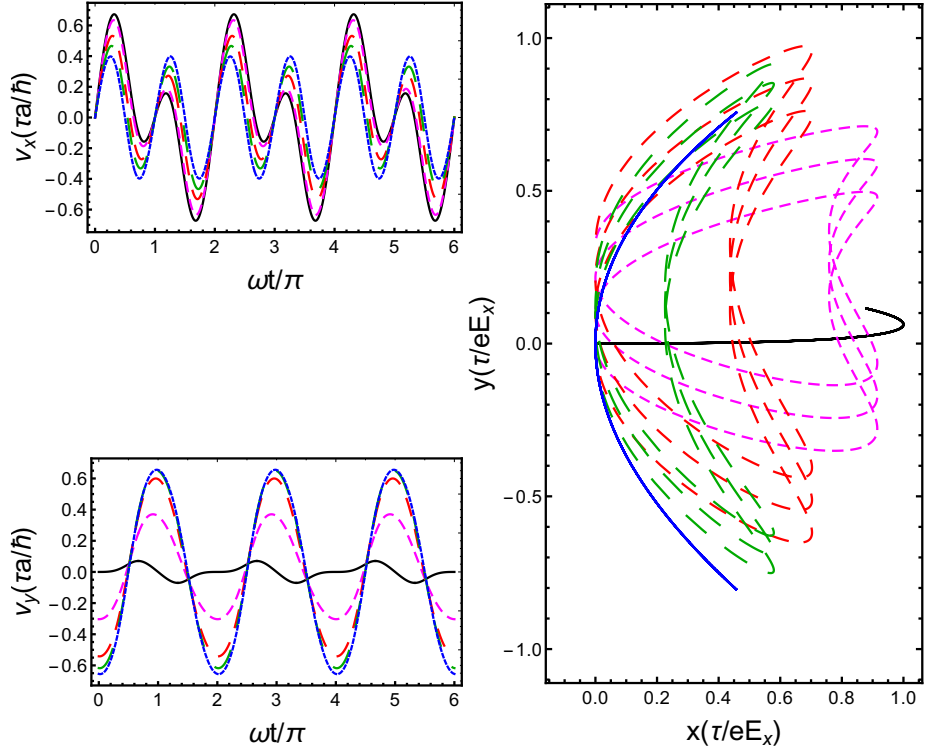}
	\caption{The time dependence of $v_x$, $v_y$ and the trajectories of the electron on the graphene sheet for different $k_y$ at $\frac{\sqrt{3}}{2}ak_x = 0$ (solid dark line), $\frac{\sqrt{3}}{2}ak_x = \pi/6$ (dash magenta line), $\frac{\sqrt{3}}{2}ak_x = \pi/3$ (dash red line), $\frac{\sqrt{3}}{2}ak_x = 5\pi/12$ (dot green line), and $\frac{\sqrt{3}}{2}ak_x = \pi/2$ (solid blue line). The electric field $\mathbf{E}$ is along the $x$ direction. }
	\label{FULLEx}
\end{figure}

Within a semiclassical framework, we start considering a static and uniform electric field $\mathbf{E}=E_x\hat{e}_x+E_y\hat{e}_y$ oriented along the graphene plane. BO are described according to the equations of motion
\begin{eqnarray}
\frac{d\mathbf{k}}{dt}&=&-e \mathbf{E}\;,\label{em1}\\
\frac{d\mathbf{r}}{dt}&=&\frac{\partial \epsilon(\mathbf{k})}{\partial \mathbf{k}}\;, \label{em2}
\end{eqnarray}
where $e$ is the quasiparticle charge, $\mathbf{r}=(x,y)$ its position vector, $\epsilon(\mathbf{k})$ is the honeycomb dispersion relation described by Eq.~(\ref{disprel}) and we assume $\hbar=1$. Combining Eqs.~(\ref{disprel}) and~(\ref{em2}), we directly obtain the semiclassical velocities along each direction
\begin{eqnarray}\label{velo}
v_x&=&-\frac{\sqrt{3}\tau a \left[
	\cos\left( \frac{3}{2}ak_x \right) 
	\sin\left( \frac{\sqrt{3}}{2}ak_y\right)
	+ 2\sin\left( \frac{\sqrt{3}}{2}ak_y\right)\cos\left( \frac{\sqrt{3}}{2}ak_y\right)
	\right] }{\varepsilon(\mathbf{k})}\;.\nonumber\\
v_y&=& -\frac{3a\tau \cos\left(\frac{\sqrt{3}}{2}ak_y \right)\sin\left(\frac{3}{2}ak_x  \right)}{\varepsilon(\mathbf{k})}\;, \label{eq:signals_vx_vy}
\end{eqnarray}
Furthermore, integrating Eq.~(\ref{em1}) we have 
\begin{eqnarray}\label{ks}
k_x(t)\ =\  k_x(0)-eE_x t\;,\qquad
k_y(t)\ =\  k_y(0)-eE_y t\;,
\end{eqnarray}
where $k_{x,y}(0)$ are the components of the initial wave vector, i.e., $\mathbf{k}_0=(k_x(0),k_y(0))$. In our discussion, we set $\mathbf{k}_0$ considering three different scenarios with the intention of comparing with results shown in \cite{Chen}:
\begin{itemize}
	\item [I.] The electric field $\mathbf{E}=E_y\hat{e}_y$ and $\mathbf{k}_0 = (0,(\pi/6)(2/\sqrt{3}a))$. \label{case1}
	\item [II.] The electric field $\mathbf{E}=E_x\hat{e}_x$ and $\mathbf{k}_0 = (0,(\pi/4)(2/\sqrt{3}a))$. \label{case2}
	\item [III.] The electric field \label{case3} $\mathbf{E}=E_x\hat{e}_x+E_y\hat{e}_y$ and $\mathbf{k}_0 = 0$.
\end{itemize}
In either case, inserting $k_{x,y}(t)$ from Eq.~(\ref{ks}) into the semi-classical velocities~(\ref{velo}), and integrating with respect to time, we obtain the trajectories for BO in graphene. These trajectories are no longer expressed in a closed form, and the integrals involved have to be solved numerically. Similar expressions were discussed in \cite{Chen}, though the typographical errors of that work were corrected in our work (see Figures \ref{FULLEy} and~\ref{FULLEx}). We generate curves corresponding to the semi-classical velocities to setup the ANN as we discussed below. 

\section{Signals creation and feature processing} \label{sec:signals}
Using Eq. (\ref{eq:signals_vx_vy}), we have simulated BO varying the electric field applied into the crystal structure, considering $\hbar=e=1$ and $\tau=a=1/2$. Therefore, the result is a time series for both electron velocities ($v_x$ and $v_y$). The signals time lapse depends on the frequency of them, since for higher frequencies, a larger sampling is required to describe the BO signals appropriately, in terms of precision and available computational resources, as we have already analyzed in \cite{els}. With this in mind, both time series ($v_x$ and $v_y$), have been discretized into 100 values each, that work as the input data for the ANN. This means, that in general, an input vector has 200 elements described by:
\begin{equation}
\mathbf{I}= \left( v_x(t_0),v_y(t_0),\ldots,v_x(t_{99}),v_y(t_{99}) \right).
\label{eq:Inputdata}
\end{equation}

On each scenario, the ANN has been trained with a supervised learning algorithm, which means that we need to specify the corresponding targets, i.e, the electric field employed to create the signals. Similarly to the work done in \cite{els}, the ANN works as a classifier. This means that the outputs of the ANN are associated with classes, defining different ranges of the electric field used to generate the BO signal. In other words, each input data is a pattern to be categorized within an electric field range, which belongs to an specific class. Next, the electric field ranges of these classes are specified for each case of study. 

Once the input data and targets for each signal have been specified, the ANN is trained using this information. The training and validation patterns set were selected randomly from the whole set of signals, where the number of signals in each set depends on the case of study, as specified on Table \ref{Table:cases}. Additionally, we evaluate the performance of the ANN using a test set with the same number of signals as the validation set, in which the patterns were built using random values of the electric field withing the predefined classes. Given that the structure of the ANN depends on the case, we discuss each scenario separately.\\

\begin{table}[h]
	\begin{center}
		\begin{tabular}{|c|c|c|c|}
			\hline 
			& Case I & Case II & Case III\tabularnewline
			\hline
			\hline 
			Training set & 350 & 350 & 1750 \tabularnewline \hline
			Validation set & 150 & 150 & 750 \tabularnewline 
			\hline
			Test set & 150 & 150 & 750 \tabularnewline 
			\hline
            \hline		
			Hidden neurons & 25 & 45 & 20 \tabularnewline 
			\hline
			Output neurons & 1 & 1 & 2 \tabularnewline 
			\hline
			Learning rate & $10^{-2}$ & 7x$10^{-3}$ & 5x$10^{-3}$ \tabularnewline 
			\hline
			Iterations & 5x$10^4$ & 5x$10^4$ & $10^4$ \tabularnewline 
			\hline
		\end{tabular}
		
		\caption{Number of patterns in each set and the parameters used in the ANN for each case of study.}
		\label{Table:cases}
	\end{center}
\end{table}

\subsection{ANN considerations and targets definition}
We have used a feedforward ANN trained with an offline supervised backpropagation algorithm \cite{ANN}. The algorithm minimizes a mean square error (MSE) function using a gradient descendent technique. At the beginning of the learning process, the weights are initialized randomly between [-1,1] and is selected the number of iterations for the learning taking into account that if the validation error increases, the learning ends. The validation error is obtained also using the MSE function but calculated over the patterns in the validation set. 

\subsubsection{Case I.}

As referred on Table \ref{Table:cases}, in this case 500 simulations were created with the purpose of training the ANN and validate its learning. These signals were generated varying the values of $E_y$ from the value
\begin{equation}
E_{y_{min}}=-\pi/(4 \sqrt{3})+\pi/(4 \sqrt{3}*500),
\end{equation}
to a maximum value of
\begin{equation}
E_{y_{max}}=\pi/(4 \sqrt{3})-\pi/(4 \sqrt{3}*500),
\end{equation}
with steps $\Delta E_y$ of 
\begin{equation}
\Delta E_y=\pi/(2 \sqrt{3}*500).
\end{equation}

\begin{figure}
	\centering
	\includegraphics[width=8cm]{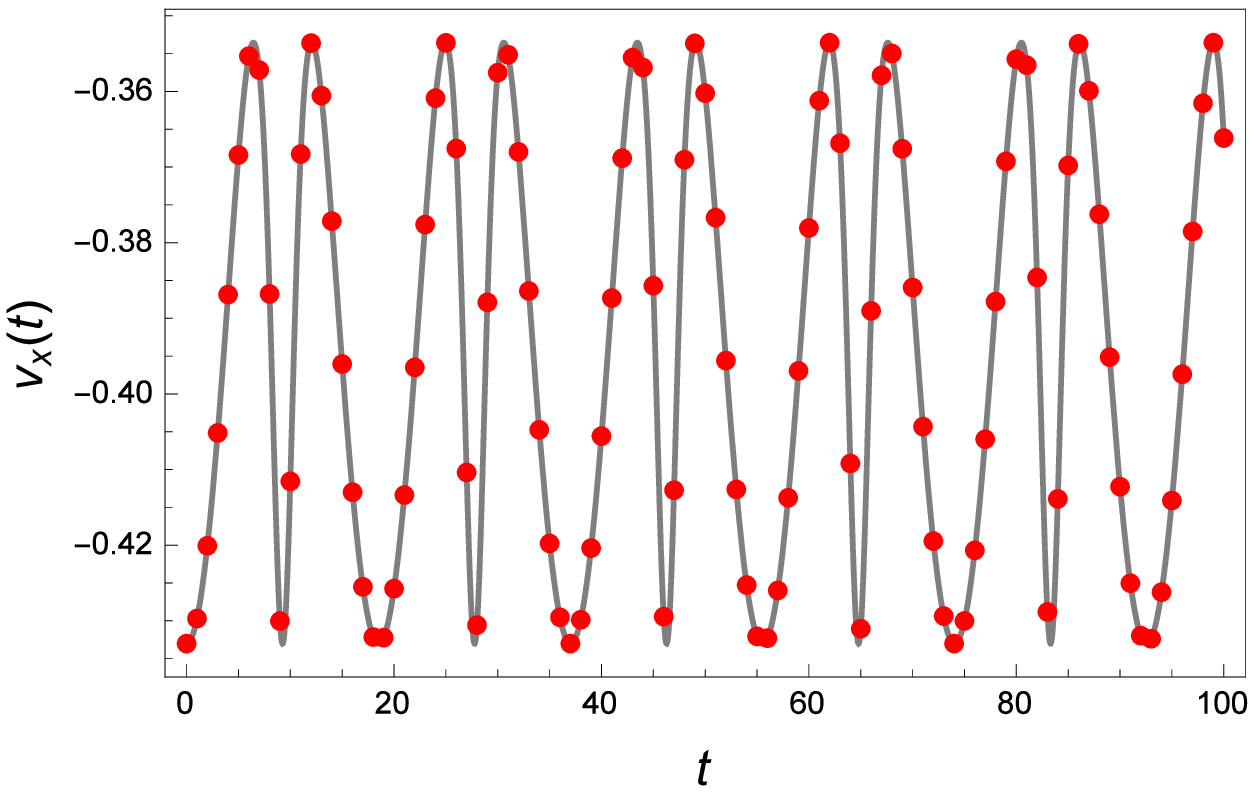}
	\includegraphics[width=8cm]{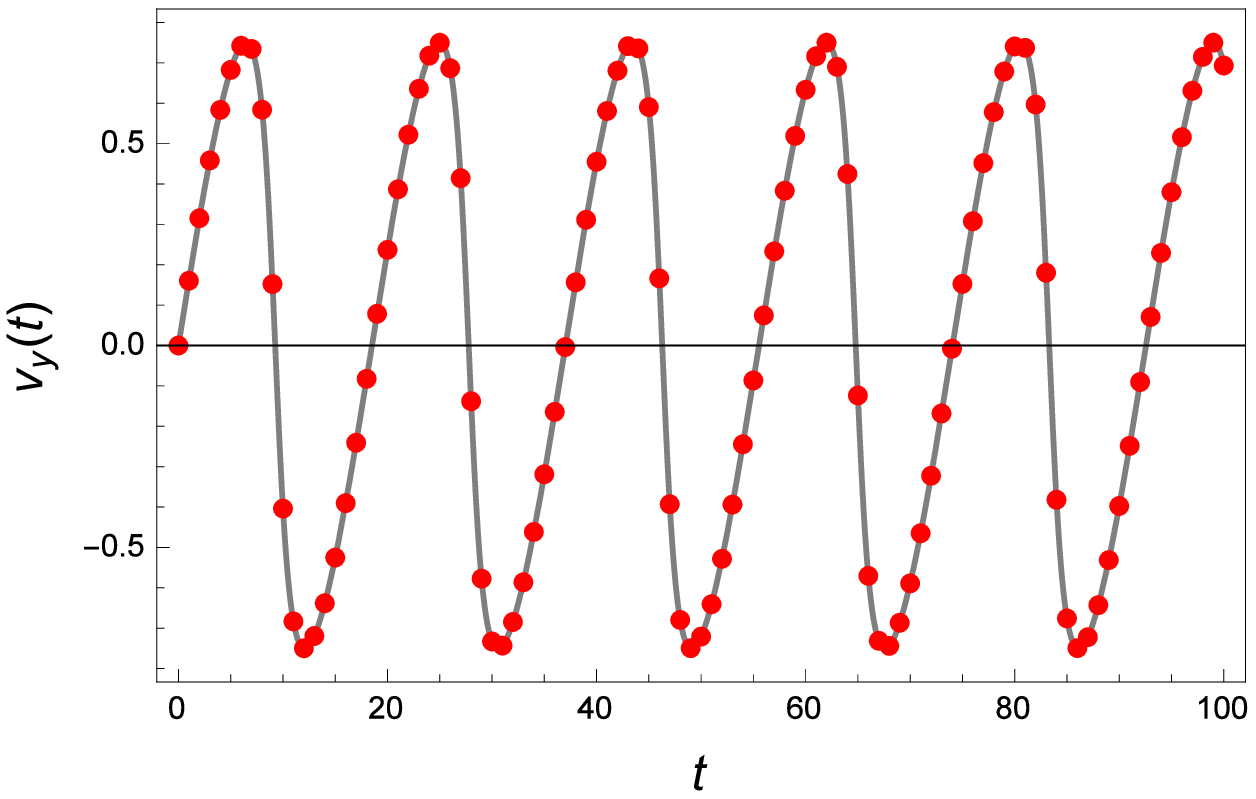}
	\caption{Sample of the velocities of the oscillating electrons generated using the considerations for the case I. The points show the values used as inputs for the ANN. Left: velocity $v_x(t)$ for $E_x=0$. Right: velocity $v_y(t)$ for $E_y=E_{y_{min}}$. }
	\label{Ex0}
\end{figure}

Given that the value of $E_x$ is equal to 0, all the time series regarding to $v_x$ are the same for the simulations in this scenario. However, we have choose to include them as input for the ANN, in terms of the general case (III). An example of a BO signal for this case is shown in Figure \ref{Ex0}.

With this considerations, we construct an ANN with a single output associated with different value ranges of $E_y$, by splitting the total of signals in subgroups or classes. For this case, we have considered 100 different classes for $E_y$. Given that we have generated 500 signal examples each class has five signals, in terms of the training and validation sets. In a mathematical representation, each class represents a range of values for $E_y$ as
\begin{equation}
Cl_m= E_{y_{min}}+\frac{5\Delta E_y}{2}m \pm \frac{5\Delta E_y}{2}; \hspace{1cm} 1 \le m \le 100,
\label{cl}
\end{equation}
with $Cl_m$ index the class number. For example, the first five values of $E_y$ correspond to the class $Cl_1$, with the center value
\begin{equation}
E_y= E_{y_{min}}+5\Delta E_y/2
\end{equation}
with an intrinsic error of $\pm 5\Delta E_y/2$. This error is equivalent to $\pm 0.05$\%, respect to the total length of the interval $[E_{y_{min}}, E_{y_{max}}]$.

Like the possible values of the sigmoid function are in the open interval (0,1), it is required to select one hundred values, one for each class in this interval. Hence, if $E_y$ correspond to the $m$ class, its corresponding target is
\begin{equation}
T=\frac{1}{200}+\frac{m-1}{100}.
\label{target}
\end{equation}
In the example considered before with $m=1$, the targets associated to the first five values of $E_y$ are $T=1/200$.

Once the training is completed, we proceeded to create a test set with the same amount of patterns as the validation set, considering random values for $E_y$ between $[-\pi/(4 \sqrt{3}), \pi/(4 \sqrt{3})]$. It is specified that a pattern is classified correctly if the output $O$ satisfies the condition
\begin{equation}
T-\frac{1}{200}< O < T+\frac{1}{200}. \label{eq:predclass}
\end{equation}

The performance of the network is measured depending on the number of patterns that are classified correctly.
\begin{figure}[h]
	\centering
	\includegraphics[width=8cm]{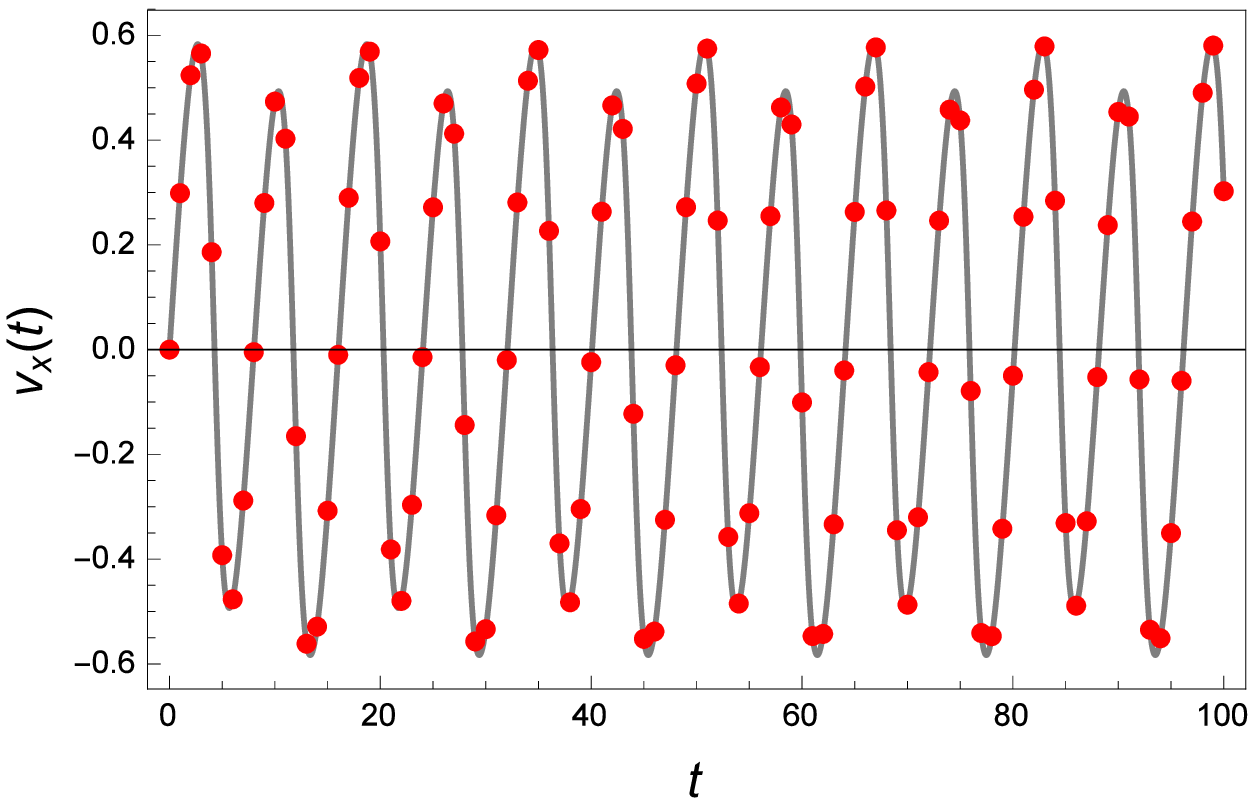}
	\includegraphics[width=8cm]{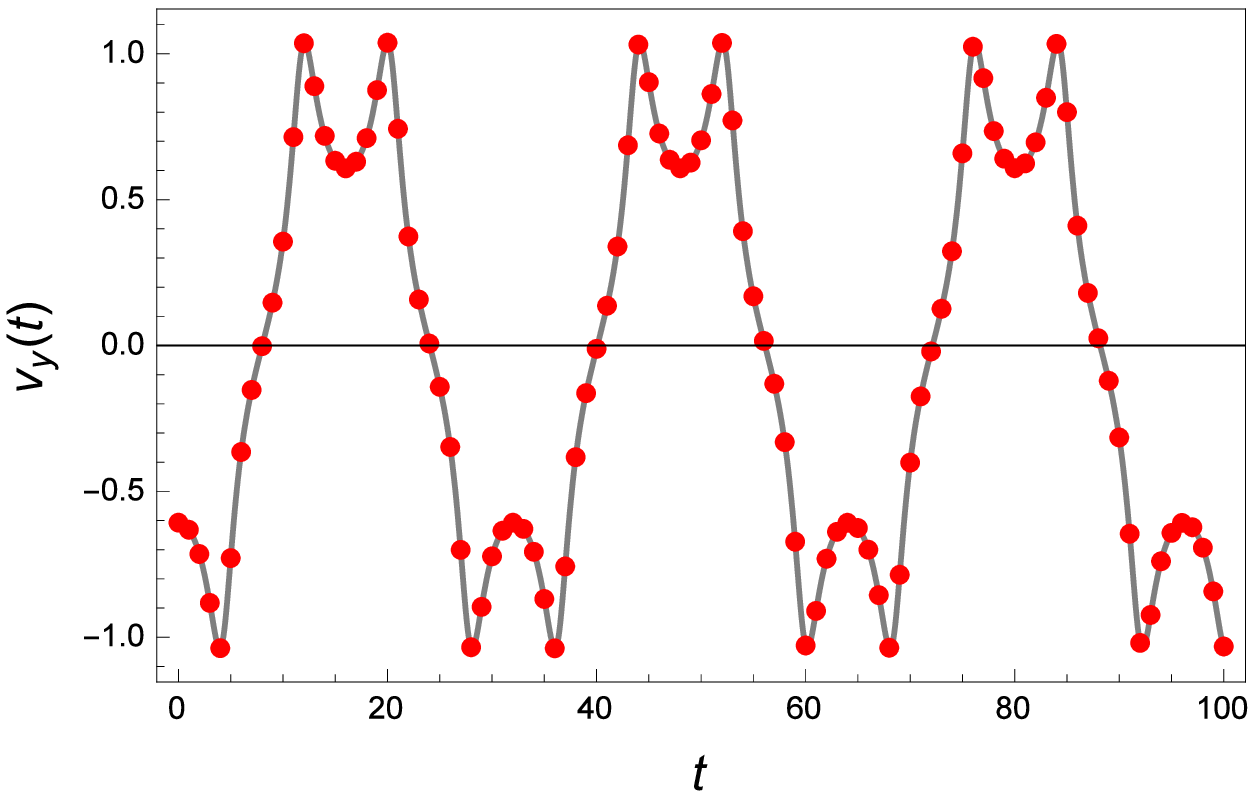}
	\caption{Sample of the velocities of the oscillating electrons generated using the considerations for the case II. The points show the values used as inputs for the ANN. Left: velocity $v_x(t)$ for $E_x=E_{x_{min}}$. Right: velocity $v_y(t)$ for $E_y=0$. }
	\label{Ey0}
\end{figure}

\subsubsection{Case II.}
The signal analysis is analogous to the previous case, but now  $E_y=0$, and $E_x$ is defined within the interval from
\begin{equation}E_{x_{min}}=-\pi/(4 \sqrt{3})+\pi/(4 \sqrt{3}*500),
\end{equation}
to
\begin{equation}
E_{x_{max}}=\pi/(4 \sqrt{3})-\pi/(4 \sqrt{3}*500),
\end{equation}
in steps of
\begin{equation}
\Delta E_x=\pi/(2 \sqrt{3}*500).
\end{equation}

A sample of a $v_x$ and $v_y$ time series for this case is shown in Figure \ref{Ey0}. The definition of classes and targets are the same as in case I, according to Eqs. (\ref{cl}) - (\ref{eq:predclass}).

\subsubsection{Case III.}
\begin{figure}
	\centering
	\includegraphics[width=8cm]{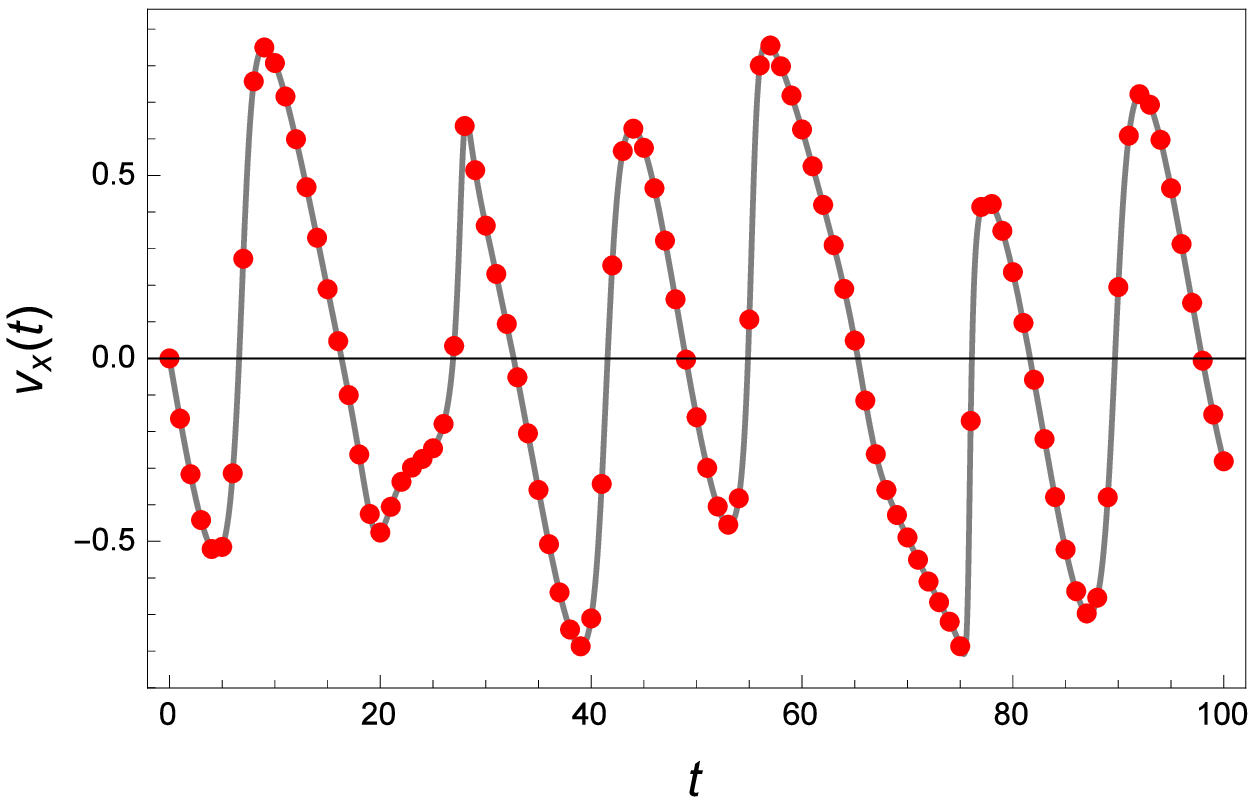}
	\includegraphics[width=8cm]{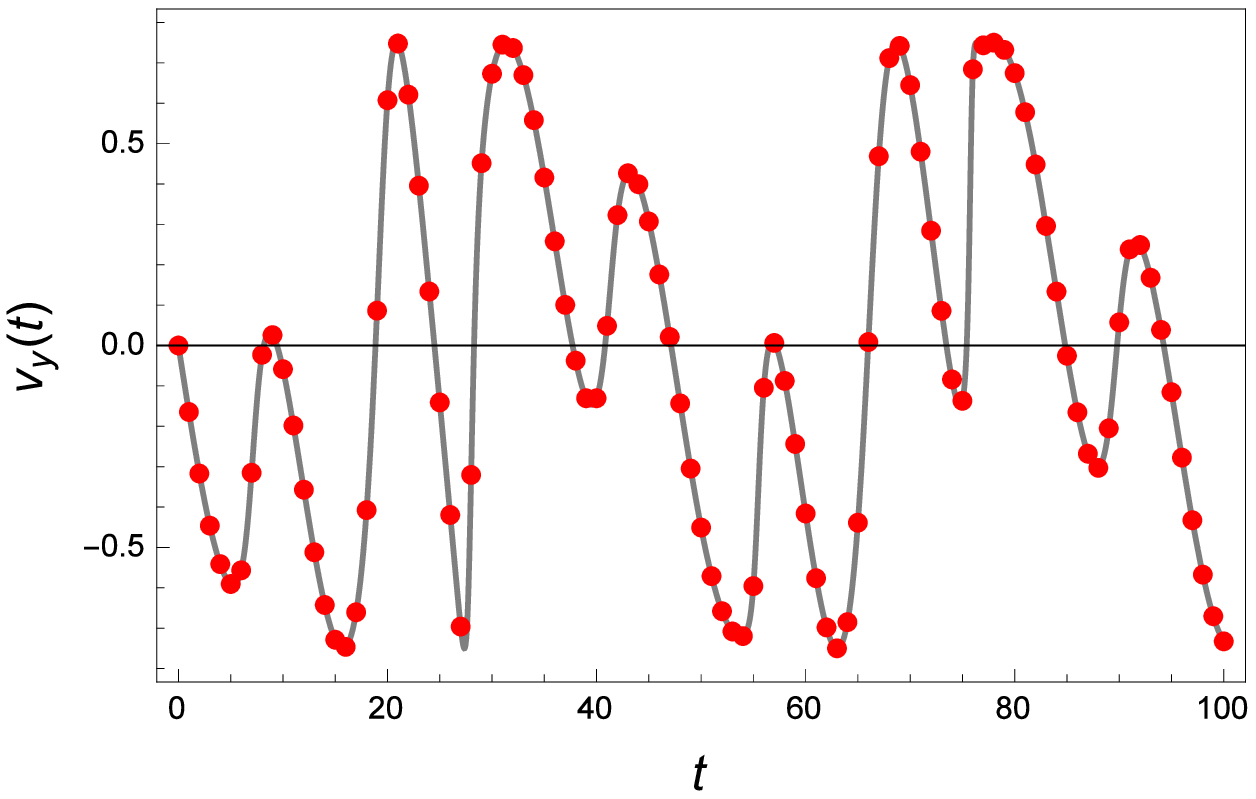}
	\caption{\label{vel} Sample of the velocities of the oscillating electrons generated using the considerations for the case III. The points show the values used as inputs for the ANN. Left: velocity $v_x(t)$ for $E_x=-\pi/(4 \sqrt{3})+\pi/(4 \sqrt{3}*50)$. Right: velocity $v_y(t)$ for $E_y=-\pi/(4 \sqrt{3})+\pi/(4 \sqrt{3}*50)$. }
	\label{grl}
\end{figure}

This is the more general scenario, where both components of the electric field were generated in steps of
\begin{equation}
\Delta E_x=\Delta E_y=\pi/(2 \sqrt{3}*50)
\end{equation}
within the interval
\begin{equation}
E_x , E_{y} \in [-\pi/(4 \sqrt{3})+\pi/(4 \sqrt{3}*50), \pi/(4 \sqrt{3})-\pi/(4 \sqrt{3}*50)].
\end{equation}
With these equations, 2500 BO signals were generated, defining 50 different values for each component of the electric field. In this case the network has 20 neurons in the hidden layer and two output neurons. The two outputs are related to each component of the electric field: the first one is related to $E_x$ and the second to $E_y$. \\
The procedure to extract the data from the created signals is the same as in the two first cases (see Figure \ref{grl}). We identify a range of values for the electric field with a class, having 10 classes for each component of the electric field
\begin{eqnarray}
Cl_{m_1}= E_{m_{1}}+\frac{5\Delta E_x}{2}{m_1} \pm \frac{5\Delta E_x}{2}; \hspace{1cm} 1 \le {m_1} \le 10\nonumber\\
Cl_{m_2}= E_{y_{m_2}}+\frac{5\Delta E_y}{2}{m_2} \pm \frac{5\Delta E_y}{2}; \hspace{1cm} 1 \le {m_2} \le 10,
\label{eq:classIII}
\end{eqnarray}
with $m_1$ and $m_2$ the classes associated to $E_x$ and $E_y$ respectively. In this case the error associated to each component of the predicted electric field is of $\pm 5$\%.\\
Therefore, the proposed targets are defined by
\begin{eqnarray}
T_1=\frac{1}{20}+\frac{m_1-1}{10}, \nonumber \\
T_2=\frac{1}{20}+\frac{m_2-1}{10},
\label{eq:targetIII}
\end{eqnarray}
where $m_1$ and $m_2$, correspond to $E_x$ and $E_y$ respectively. 
\\
Finally for each pattern, it is specified that the outputs are classified correctly by the ANN if the condition
\begin{equation}
T_{\kappa}-\frac{1}{20}< O_{\kappa} < T_{\kappa}+\frac{1}{20}
\end{equation}
is satisfied for $1 \le \kappa \le 2$. Below are the results of the training for each of the cases.

\section{Results}\label{sec:results}

As stated in Eq. (\ref{eq:Inputdata}), in all three cases the ANN consists of an input layer with 200 neurons, and an output layer as defined on each of the cases. However, the number of hidden neurons was chosen based on an ANN configuration that lead to the best performance in the ANN, this is, the ANN with the minimum error in the validation set during the learning phase. All the neurons in the hidden and output layer have a sigmoid activation function.\\
The selection of the best ANN was made by varying the number of neurons from five to fifty in steps of five. Likewise, the learning rate was chosen by exploring values within the range [$10^{-3},10^{-2}$] in steps of $10^{-3}$. The ANN specifications regarding each case of study are specified also in Table \ref{Table:cases}. The performance of the best ANN for each case is reported in the Table \ref{Table} with the percentage of patterns classified correctly (PPCC) from its corresponding set.

\begin{table}[h]
	\begin{center}
		\begin{tabular}{|c|c|c|c|c|c|c|}
			\hline 
			& \multicolumn{2}{c|}{Training set} & \multicolumn{2}{c|}{Validation set} & \multicolumn{2}{c|}{Test set}\tabularnewline
			\hline
			\hline 
			\multicolumn{7}{|c|}{Case I}\\
			\hline
			Output & \multicolumn{2}{c|}{$O_1$} & \multicolumn{2}{c|}{$O_1$}& \multicolumn{2}{c|}{$O_1$}\tabularnewline
			\hline 
			PPCC (\%) & \multicolumn{2}{c|}{94}&\multicolumn{2}{c|}{71.3} &\multicolumn{2}{c|}{84.6}\tabularnewline
			\hline 
			\hline
			\multicolumn{7}{|c|}{Case II}\\
			\hline
			Output & \multicolumn{2}{c|}{$O_1$} & \multicolumn{2}{c|}{$O_1$}& \multicolumn{2}{c|}{$O_1$}\tabularnewline
			\hline 
			PPCC (\%) & \multicolumn{2}{c|}{97.7}&\multicolumn{2}{c|}{56} &\multicolumn{2}{c|}{82.6}\tabularnewline
			\hline 
			\hline
			\multicolumn{7}{|c|}{Case III}\\
			\hline
			Output & ${O}_{1}$ & ${O}_{2}$ & ${O}_{1}$ & ${O}_{2}$ & ${O}_{1}$ & ${O}_{2}$\tabularnewline
			\hline 
			PPCC (\%) & 97.7 & 99 & 92.6 & 92.5 & 91.3 & 87.8\tabularnewline
			\hline 
		\end{tabular}
		
		\caption{PPCC by the ANN for the training, validation and test set in each case.}
		\label{Table}
	\end{center}
\end{table}
In case I, the best ANN structure was obtained with 25 hidden neurons. With this ANN, an 94\% and 71.3\% of the patterns were classified correctly from the training and validation set after 50000 iterations with a learning constant $\gamma = 1\times10^{-2}$. Meanwhile, in the test the ANN achieved a $\text{PPCC}  =  84.6\%$.\\
In the second scenario, the selected ANN with 45 hidden neurons, achieved an 97.7\% and 56\% of PPCC for the training and validations sets respectively. In the prediction set, the ANN diminished its performance a 2\% respect to case I.\\
Finally for case III, after the selection process with 10000 iterations in the training phase with learning rate $\gamma = 5\times10^{-3}$, the ANN with the best performance was the one with 20 hidden neurons. Taking into account Eqs. (\ref{eq:classIII}) and (\ref{eq:targetIII}), the ANN got for the training set a PPCC of 97.7\% and 99\% for output 1 and 2, respectively. Meanwhile, in the validation set, the ANN classified correctly the 92.6\% and 92.5\% of patterns corresponding to the $E_x$ and $E_y$ classes respectively. Finally, testing the ANN with complete unknown patterns, it got a PPCC o 91.3\% for the first output, and a 87.8\% for the second output.\\
\begin{figure}
	\centering
	\includegraphics[width=8cm]{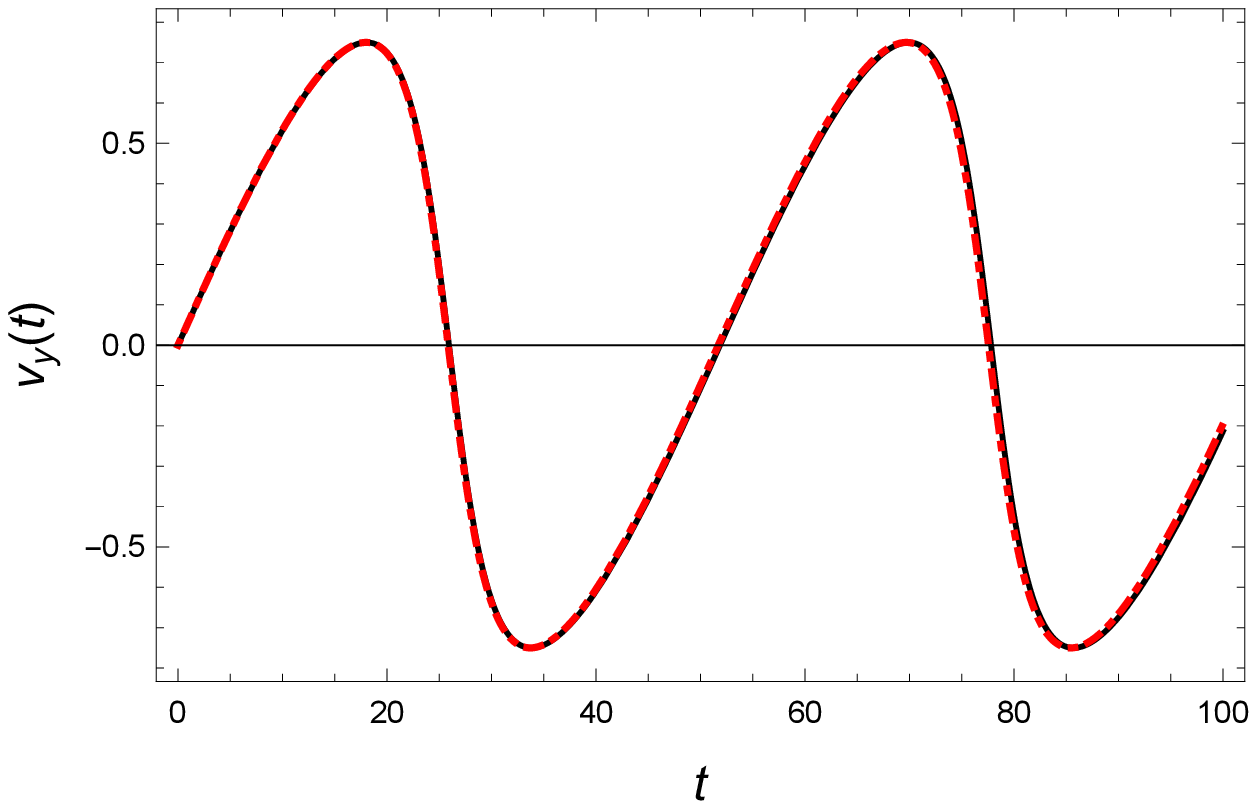}
	\includegraphics[width=8cm]{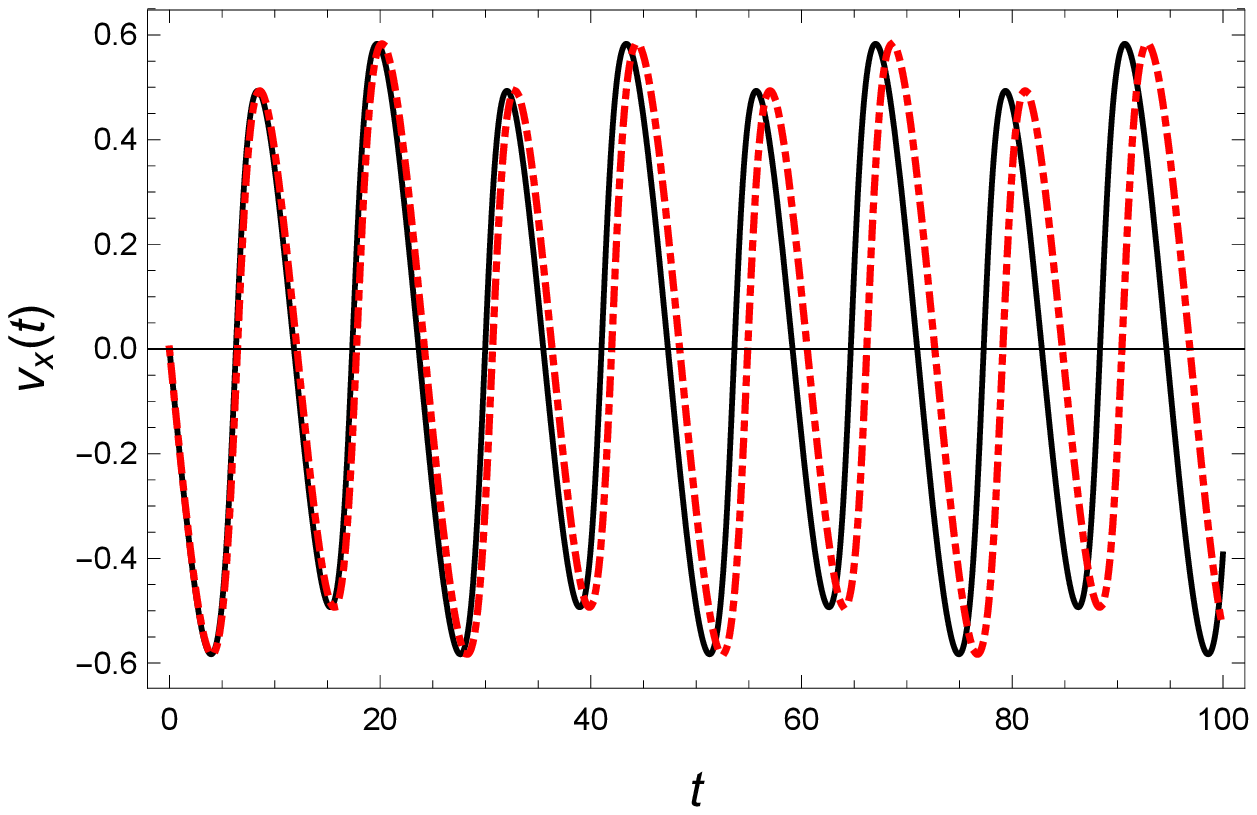}
	\caption{Left: velocity of an electron using the conditions for the case I with a random $E_y$ (continuous line) and the electron velocity using the value $E_y$ predicted by the ANN (Dashed line) with an error of $\pm0.5\%$. Right: velocity of an electron using the conditions for the case II with a random $E_x$ (continuous line) and the electron velocity using the value $E_x$ predicted by the ANN (Dashed line) with an error of $\pm1\%$. }
	\label{pre}
\end{figure}
\\
It can be seen that for cases I and II the performance of the ANN is not good as for case III, but this is because the first cases have more classes and the error associated to the predicted electric field is less. Meanwhile, for the last case, although the PPCC is larger, the uncertainty associated to the predicted electric field target also is larger. For this reason, if is considered a larger interval for the output as a correct classification
\begin{equation} 
T-\frac{1}{100}< O < T+\frac{1}{100},
\end{equation}
where the output is associated to each component of the electric field depending on the case, the PPCC increase but having a larger error in the predicted electric field for cases I and II, as it can be seen in Figure \ref{pre}. The performance of the ANN taking this consideration is shown in the Table \ref{Table2}.\\

\begin{figure}
	\centering
	\includegraphics[width=8cm]{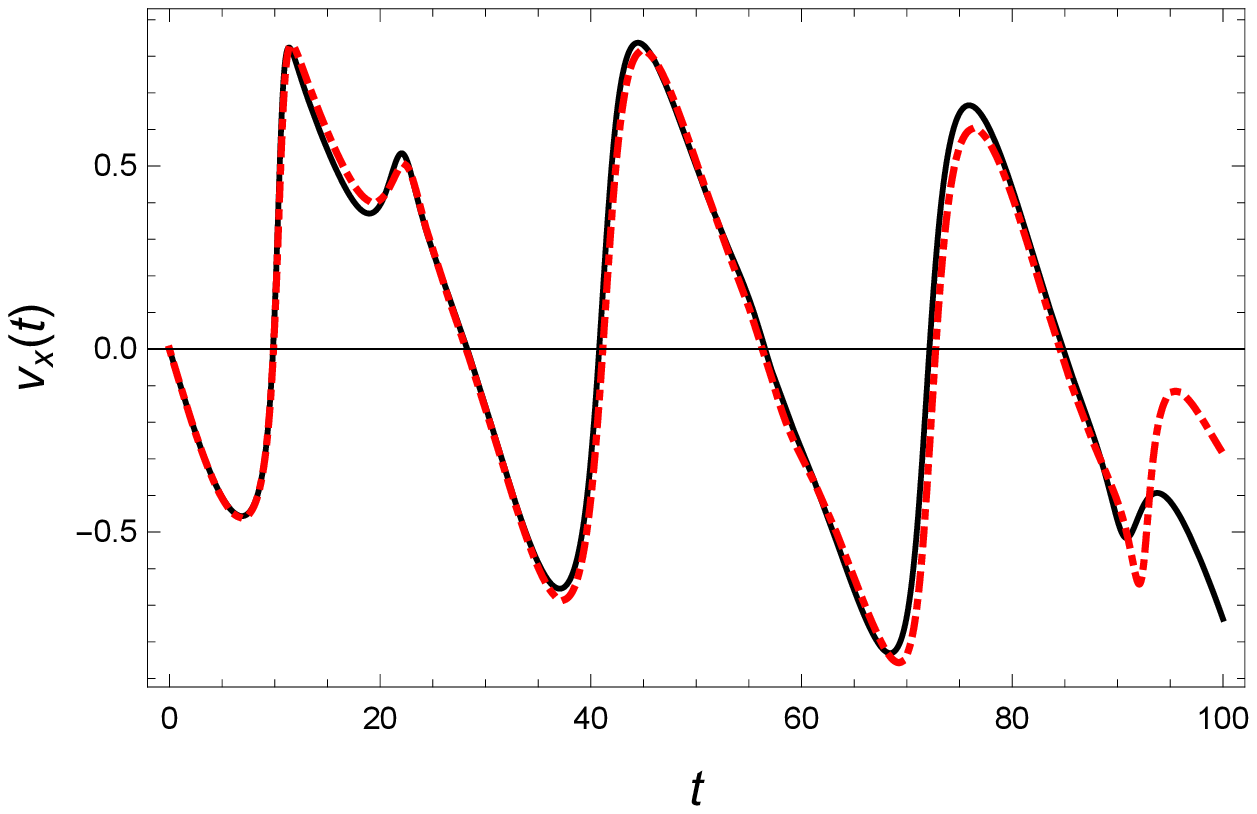}
	\includegraphics[width=8cm]{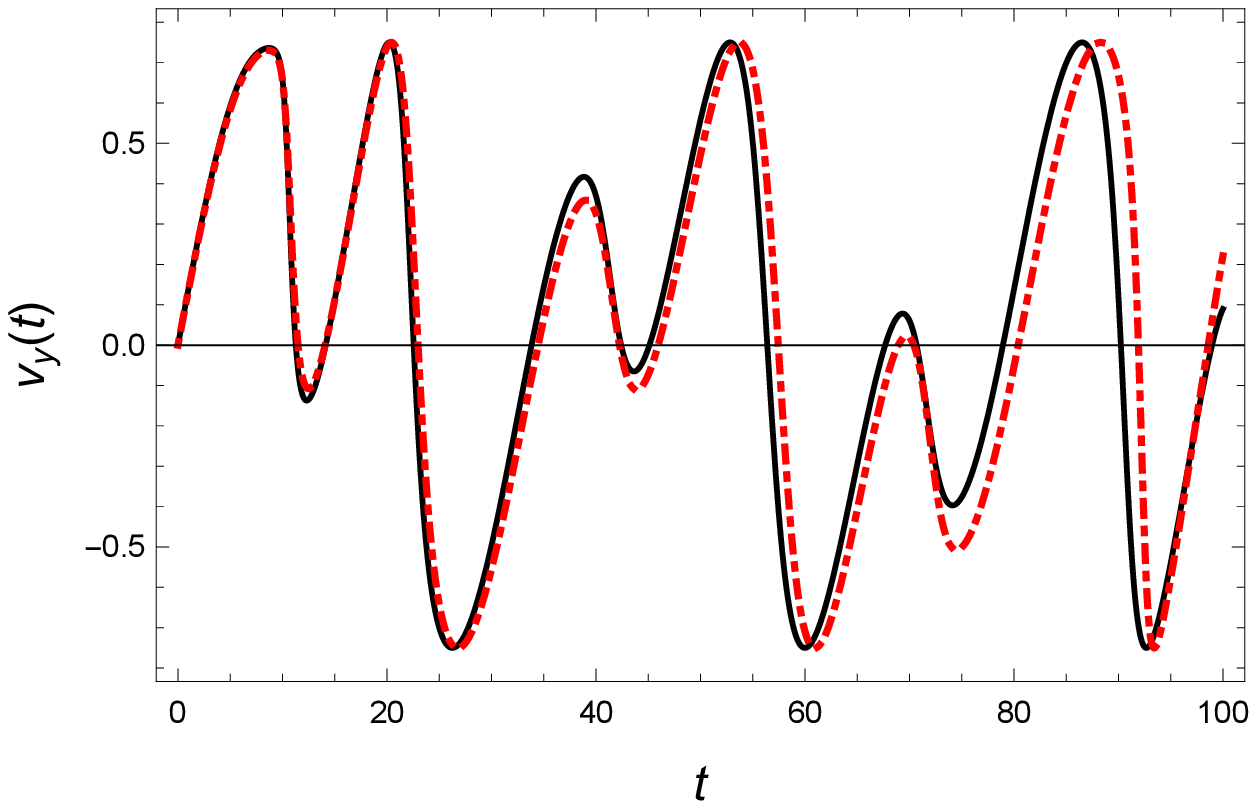}
	\caption{Velocities of an electron using the conditions for the case III with a random $E_x$ and $E_y$ (continuous lines) and the electron velocities using the values $E_x$ and $E_y$ predicted by the ANN (Dashed lines) with high accuracy.}
	\label{pregrl}
\end{figure}

\begin{figure}
	\centering
	\includegraphics[width=8cm]{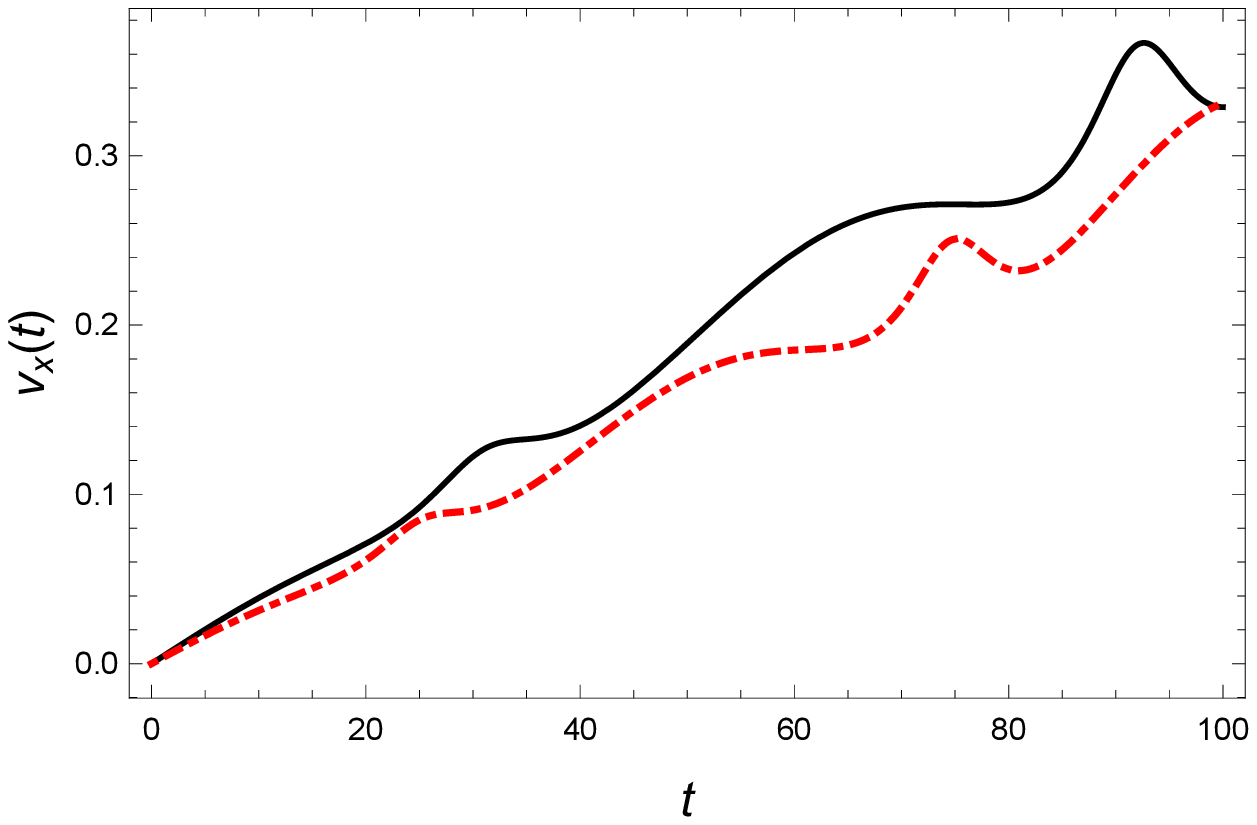}
	\includegraphics[width=8cm]{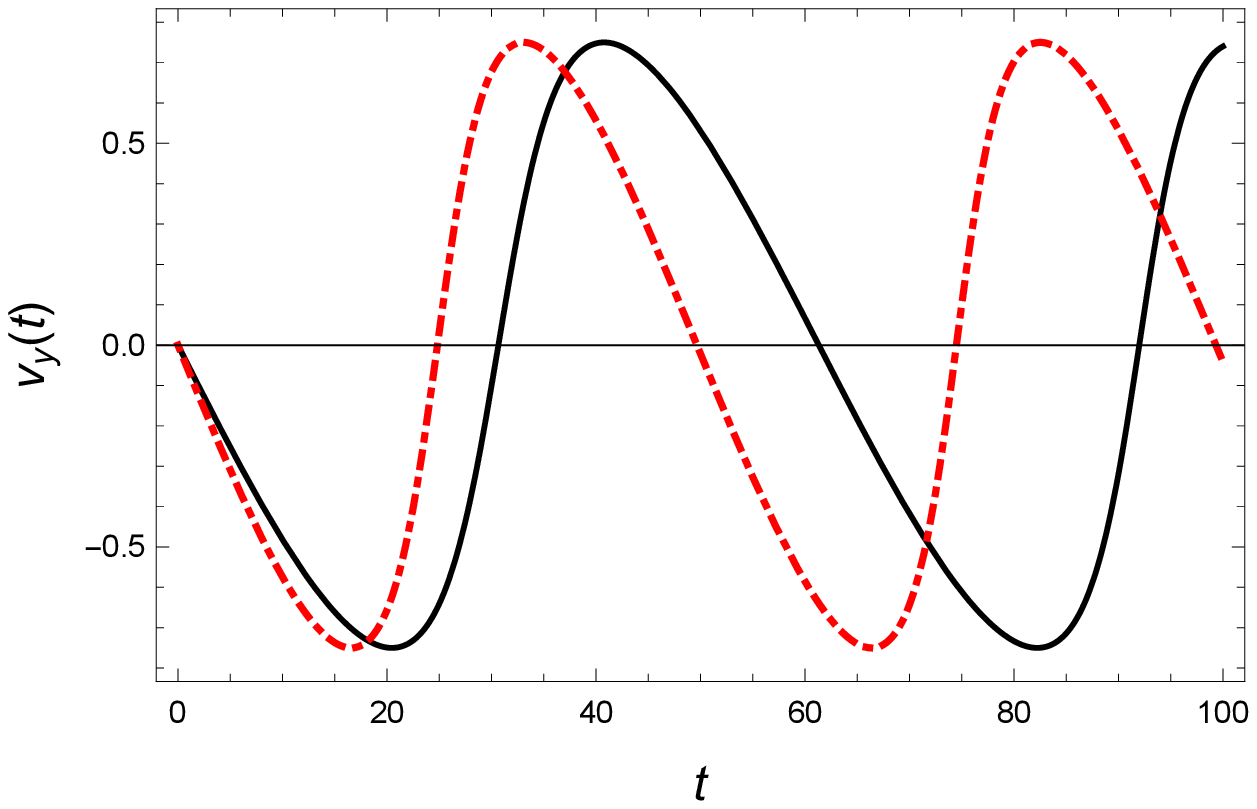}
	\caption{Velocities of an electron using the conditions for the case III with a random $E_x$ and $E_y$ (continuous lines) and the electron velocities using the values $E_x$ and $E_y$ predicted by the ANN (Dashed lines) with low accuracy.}
	\label{pregrl1}
\end{figure}

\begin{table}[h]
	\begin{center}
		\begin{tabular}{|c|c|c|c|c|c|c|}
			\hline 
			& \multicolumn{2}{c|}{Training set} & \multicolumn{2}{c|}{Validation set} & \multicolumn{2}{c|}{Test set}\tabularnewline
			\hline
			\hline 
			\multicolumn{7}{|c|}{Case I}\\
			\hline
			Output & \multicolumn{2}{c|}{$O_1$} & \multicolumn{2}{c|}{$O_1$}& \multicolumn{2}{c|}{$O_1$}\tabularnewline
			\hline 
			PPCC (\%) & \multicolumn{2}{c|}{100}&\multicolumn{2}{c|}{98} &\multicolumn{2}{c|}{99.3}\tabularnewline
			\hline 
			\hline
			\multicolumn{7}{|c|}{Case II}\\
			\hline
			Output & \multicolumn{2}{c|}{$O_1$} & \multicolumn{2}{c|}{$O_1$}& \multicolumn{2}{c|}{$O_1$}\tabularnewline
			\hline 
			PPCC (\%) & \multicolumn{2}{c|}{100}&\multicolumn{2}{c|}{84.6} &\multicolumn{2}{c|}{94.6}\tabularnewline
			\hline 
		\end{tabular}
		\caption{PPCC by the ANN considering a greater error in the predicted electric field for cases I and II.}
		\label{Table2}
	\end{center}
\end{table}
A sample of an electron velocity constructed using the electric field component predicted by the ANN for case III is shown in Figures \ref{pregrl} and \ref{pregrl1}. In these Figures we can see that sometimes the predicted velocities are accurate (Figure \ref{pregrl}), but due the error of $\pm 5\%$ in the predicted electric field, it can happen that the predicted velocity differs from the real even when it was classified correctly (Figure \ref{pregrl1}).

\section{Final remarks}\label{sec:conclusions}
The developed ANN classify numerical signals corresponding to BO in three situations depending on the electric field involved. \\

In the first case, the trained ANN classify signals corresponding to a BO where $E_y \in [-\pi/(4\sqrt{3}),\pi/(4\sqrt{3})]$ and $E_x = 0$ for a initial momentum $k_0$ constant. An 84.6\% of the signals are classified correctly within an error of $\pm 0.5$\% for the predicted $E_y$. When is considered an error of $\pm1$\%, the percentage of correct classifications rises to a $99.3$\%.

For the second case, the trained ANN classify signals corresponding to a BO where $E_x \in [-\pi/(4\sqrt{3}),\pi/(4\sqrt{3})]$ and $E_y=0$ with an initial momentum constant. A 82.6\% of the signals are classified correctly with an error of $\pm 0.5$\% for the predicted $E_x$. When is considered an error of $\pm1$\%, the percentage of correct classifications rises to a $94.6$\%.

In the last case, the trained ANN classify signals corresponding to a BO where both components of the electric field are between the interval $[-\pi/(4\sqrt{3}),\pi/(4\sqrt{3})]$ for a fixed initial momentum $k_0$. The ANN classify correctly 91.3\% the component $E_x$ of the signals with an error of $\pm5$\%, meanwhile the component $E_y$ is classified correctly 87.8\% also with an error of $\pm5$\%.

As a natural extension of the work, we are considering the influence of strain in graphene on Bloch Oscillations from the Artificial Neural Networks perspective as a guide to look for experimental observables. Results will be reported elsewhere.

\section*{Acknowledgments}
We acknowledge support from CONACyT grant 256494 and CIC-UMSNH (M\'exico)
under grants 4.22 and 4.23. We also thank for providing computer resources to ABACUS Laboratorio de Matem\'aticas Aplicadas y C\'omputo de Alto Rendimiento del CINVESTAV-IPN under grant CONACT-EDOMEX-2011-C01-165873.

\section*{References}
\bibliographystyle{elsarticle-num}

\begin{thebibliography}{00}   
	\bibitem{Wal} Wallace, P. R.: The Band Theory of Graphite, {\it Phys. Rev.} {\bf 71} 622634 (1947).
		
			\bibitem{graphene1} Novoselov, K. S., et al.: Two-dimensional atomic crystals, {\it Proc. Natl Acad. Sci. USA} {\bf 102}, 10451 (2005).
		%
		\bibitem{graphene2}  Zhang Y, et. al.: Experimental observation of the quantum Hall effect and Berry's phase in graphene, {\it Nature} 438, 201 (2005).
		%
		\bibitem{graphene3} Geim, A. K. y Novoselov, K. S.: The rise of graphene, {\it Nature Materials} {\bf 6}, 183191 (2007).
		%
		\bibitem{Weyl} Vafek, O. y Vishwanath, A.: Dirac Fermions in Solids - from High $T_{c}$ cuprates and Graphene to Topological Insulators and Weyl Semimetals, {\it Ann. Rev. Cond. Mat. Phys}{\bf 5}; 83-112 (2014).
		%
		\bibitem{Bloch} Esaki, L. y Tsu R.: Superlattice and Negative Differential Conductivity in Semiconductors, {\it J. Res. Dev.} {\bf 61} 61 (1970).
		
		\bibitem{Zener} Zener C.:  Non-adiabatic crossing of energy levels. Proc R. Soc. A 137:696 (1932).
		
		
		\bibitem{exp1} Feldmann J., Leo K., Shah J., Miller D.~A.~B., Cunningham J.~E., Meier T., von Plessen G., Schulze A., Thomas P., and Schmitt-Rink S.: Optical investigation of Bloch oscillations in a semiconductor superlattice, {\it Phys. Rev. B}{\bf 46} 7252 (1992).
		%
		\bibitem{exp2}von Plessen G. and Thomas P.: Method for observing Bloch oscillations in the time domain, {\it Phys. Rev. B}{\bf 45}, 9185 (1992).
		
		\bibitem{exp21}Leo K., Bolivar P.~H., Br\"uggemann F., Schwedler R., and K\"ohler K.: Observation of Bloch oscillations in a semiconductor superlattice, {\it Solid State Commun.} {\bf 84}, 943 (1992).
		
		\bibitem{exp22}Leisching P., Haring Bolivar P., Beck. W., Dhaibi Y., Br\"uggemann F., Schwedler R., Kurz H., Leo K., and. K\"ohler K.: Bloch oscillations of excitonic wave packets in semiconductor superlattices {\it Phys. Rev. B}{\bf 50} 14389 (1994).
		%
		\bibitem{exp3} Dekorsy T., Leisching P., K\"ohler K., and Kurz H.: Electro-optic detection of Bloch oscillations, {\it Phys. Rev. B}{\bf 50} 8106 (1994).
		%
		\bibitem{exp31}Dekorsy T., Ott R., Kurz H., and K\"ohler K.: Bloch oscillations at room temperature, {\it Phys. Rev. B}{\bf 51} 17275 (1995).
		%
		\bibitem{exp4}	Waschke C.,  Roskos H.~G., Schwedler R., Leo K., Kurz H., and  K\"ohler K.: Coherent submillimeter-wave emission from Bloch oscillations in a semiconductor superlattice, {\it Phys. Rev. Lett.} {\bf 70}, 3319 (1993).
		\bibitem{exp41}Roskos H.~G., Waschke C., Schwedler R., Leisching P.,  Dhaibi Y., Kurz H., and K\"ohler K. Bloch oscillations in GaAs/AlGaAs superlattices after excitation well above the bandgap, {\it Superlattices and Microstructures} {\bf 15} 281 (1994).
		%
		\bibitem{exp5}Kolovsky A.~R. and  Korsch H.~J.: Bloch Oscillations in cold atoms in two-dimensional optical lattices, {\it Phys. Rev. A.} {\bf 67}, 063601 (2003).
		%
		\bibitem{exp51}Witthaut D., Keck F., Korsch H.~J. and Mossmann S.: Bloch Oscillations in two-dimensional lattices, {\it New J. Phys.} {\bf 6}, 41 (2004).

\bibitem{atom1} Dahan M. B, Peik E., Reichel J, Castin Y. and Salomon,
C. {\it Phys. Rev. Lett.} 76, 45084511 (1996).

\bibitem{atom2} Genske, M. et al., {\it Phys. Rev. Lett.} 110, 190601. (2013).

\bibitem{dielec1}  Pertsch T, Dannberg P, Elflein W, Br\"auer, A and Lederer
F 1999, {\it Phys. Rev. Lett.} {\bf 83} 4752.

\bibitem{dielec2} Morandotti R., Peschel U., Aitchison J. S., Eisenberg H. S., and Silberberg Y 1999,{\it Phys. Rev. Lett}.83, 47564759.

\bibitem{dielec3} Sapienza R et al., 2003, {\it Phys. Rev. Lett}.91, 263902.

\bibitem{plas} Block A et al, 2014, {\it Nat}. Commun. 53843.


\bibitem{cheng} Cheng Hemeng et al., Electronic Bloch oscillation in bilayer graphene gradient superlattices, Applied Physics Letters 105, 072103 (2014); doi: 10.1063/1.4893598.

\bibitem{changan} Changan Li et al, Electronic band gaps and transport properties in aperiodic bilayer graphene superlattices of Thue-Morse sequence, Appl. Phys. Lett. 103, 172106 (2013); doi: http://dx.doi.org/10.1063/1.4826643
		
		\bibitem{IBP} Gonz\'alez J.~A., Hern\'andez-Ortiz S., L\'opez C.~E. and Raya A. Bloch oscillations: Inverse problem, {\it Plasmonics} (2016) doi:10.1007/s11468-016-0477-x
		%
		\bibitem{els} M. Carrillo, Gonz\'alez J.~A., Hern\'andez-Ortiz S., L\'opez C.~E. and Raya A.: Bloch oscillations in two-dimensional crystals: Inverse problem. {\it Computational Materials Science} (2017) doi:10.1016/j.commatsci.2017.04.030
		%
		\bibitem{Chen} Chen R, Ma T, Wang L-G and Lin H-Q 2013, {\it Electronic Bloch oscillation in a pristine monolayer graphene}, e-print: arXiv:1301.3221 [cond-mat.mes-hall]		
		
		\bibitem{ANN} Rojas R. Neural Networks. A Systematic Introduction 1996 Springer-Verlag
		
	
		

	\end{thebibliography}

\end{document}